\documentclass[preprint,prd,nofootinbib,tightenlines,amsmath]{revtex4}
\usepackage{graphicx}
\usepackage{bm}
\usepackage{color}
\usepackage{subfigure}

\begin{document}
\baselineskip=15pt \parskip=5pt

\vspace*{3em}

\title{Dispersion Relations Explaining OPERA Data \\From Deformed Lorentz Transformation}

\author{Gang Guo$^1$, Xiao-Gang He$^{1,2}$ }

\affiliation{
$^1$INPAC, Department of Physics, Shanghai Jiao Tong University, Shanghai, China\\
$^2$Department of Physics and  Center for Theoretical Sciences, \\
National Taiwan University, Taipei, Taiwan
}

\date{\today $\vphantom{\bigg|_{\bigg|}^|}$}

\begin{abstract}
OPERA collaboration has reported evidence of superluminal phenomenon for neutrinos.
One of the possible ways to explain the superluminality is to have Lorentz symmetry violated.
It has been shown that
dispersion relations put forwards has the problem of physics laws vary in different inertial
frames leading to conflicting results on Cherenkov-like radiation, pion decay and high energy neutrino cosmic ray.
For theories with deformed Lorentz symmetry, by modifying conservation laws corresponding to energy and momentum in the usual
Lorentz invariant theory, it is possible to
have superluminal effect and at the same time avoid to have conflicts encountered in Lorentz violating theories.
We study dispersion relations from deformed Lorentz symmetry. We find that it is possible
to have dispersion relations which can be consistent with data on neutrinos. We show that once
the superluminality $\delta v$ as a function of energy is known the corresponding dispersion relation in the deformed Lorentz symmetry can be determined.
\end{abstract}

\maketitle

The OPERA collaboration has reported evidence of superluminal behavior for muon neutrino $\nu_\mu$ with energies of a few tens of GeV \cite{opera}. More recently,
they have reported new data with a shorter beam bunch allowing the measurement of the neutrino flight time more accurately. They confirmed their earlier results \cite{opera}. The mean neutrino energy
is 17 GeV and neutrinos arrived earlier by $\delta t = (57.8\pm 7.8(stat.)^{+8.3}_{-5.9}(sys.))$ ns than expected if neutrinos are traveling with the speed of light $c$.
This implies a superluminal velocity $v$ of neutrinos by relative amount of superluminality $\delta v = (v-c)/c = (2.37\pm 0.32(stat.)^{+0.34}_{-0.24})\times 10^{-5}$.
This has generated extensive discussions.
Existence of superluminal phenomena as large as OPERA observed challenges the completeness of relativity theory based on Lorentz symmetry~\cite{elles}.
It is important that the OPERA result
be tested by other experiments. Many of the studies in the literature aim to study theoretical explanations and implications of the data. Different mechanisms have been proposed to explain the data.
Lorentz violation is a popular scenario\cite{lorentz,strumia,tachyon}.

It has been shown that explanation of the OPERA data based on certain Lorentz violating theory is inconsistent with other known results from the same experiment \cite{cg, bi}.
Cohen and Glashow argued that in the energy range of
neutrino, the beam would loose much of their energy, via Chereov-like processes, such as $\nu_\mu \to \nu_\mu + e^+ + e^-$, on their way from CERN to the OPERA detector at Gran Sasso, and OPERA would not detect neutrinos with energy exceeding 12 GeV in certain Lorentz violating theories. Bi et al and Gonzalez-Mestres showed\cite{bi} that with the large superluminal effect of $\delta v = 2.37\times 10^{-5}$, neutrinos cannot have energies exceeding 5 GeV if they are from $\pi^+ \to \mu^+ \nu_\mu$ which is believed to provide the substantial amount of the muon neutrino flux detected at Gran Sasso. It is interesting to note that physics laws are altered in different inertial frames. In the $\nu_\mu \to \nu_\mu + e^++e^-$ case, in an inertial frame in which the muon neutrino has less than
two times of the electro rest mass, the processes cannot happen, but in another inertial frame the initial neutrino travels with a larger velocity and therefore a higher energy than two rest electron mass, this process can happen. In the $\pi^+\to \mu^+ \nu_\mu$ case, in an inertial frame pion has low energy, the process can happen, but with higher energy, it may not happen. This, by itself, is an very interesting result. This is in contrary to our common believes and needs an explanation. This conclusion applies to many Lorentz violating theories.

Several solutions to these conflicts have been proposed. If neutrinos are Tachyons, they always have a velocity larger than the speed of light and is consistent with Lorentz symmetry. However, it has been shown that this type of theories has problems when taking into account other available data\cite{tachyon}, such as possible dependence of neutrino maximal speed with energy. Another interesting possibility of explaining the data by keeping the principle of relativity of inertial frames is based on deformed Lorentz symmetry\cite{deform1,deform2,deform3,deform4}. This type of theory allows superluminal effect but can avoid the problems above by keeping physical laws not changed in different inertial frames, if the conservation laws corresponding to the energy and momentum in the usual Lorentz symmetry are suitably modified. In this work, we will work in the framework of deformed Lorentz symmetry to find dispersion relations which can have superluminal effect to explain the OPERA data and also other related neutrino data including those from neutrino oscillation.

A general dispersion relation between energy $E$ and momentum $p$ can be written as\cite{deform1}, $E^2 f^2(E,p) - p^2 g^2(E,p) = m^2$, which can also be re-written without loss of generality as
\begin{eqnarray}
F^2(E, p) E^2  - p^2 = m^2\;. \label{gg}
\end{eqnarray}
$F^2(E, p) = f^2(E,p) - (g^2(E,p)-1) p^2/E^2$. One can also solve for $E$ as a function of $p$ first and then insert the expression for $E$ in $F(E,p)$ to transform $F$
as a function of $p$ and $m$. We will write $F$ as $F(p,m)$.

If the generator of Lorentz boost $N_i$ is modified in such a way that
\begin{eqnarray}
[N_i, F^2(p,m) E^2 - p^2] = 0,
\end{eqnarray}
then the relativity of the inertial frames are preserved. In the limit of usual Lorentz symmetry $F(p,m) = 1$. If one also wants to keep the meaning of mass as rest energy, then
$F(0,m) = 1$.
The above commutation relation can be achieved by\cite{deform3}
\begin{eqnarray}
[N_i, p_j] = F(p,m) E \delta_{ij}\;,\;\;[N_i, E] = p_i\left ({1\over F(p,m)} - 2 E^2 {\partial F(p,m)\over \partial p^2}\right )\;.
\end{eqnarray}

The above defines a deformed Lorentz symmetry. With this deformed Lorentz symmetry, the conserved quantities corresponding to the energy momentum
conservation laws in the usual Lorentz symmetry is modified. They are
\begin{eqnarray}
F(p,m) E\;,\;\;\;\vec p\;.
\end{eqnarray}

It has been shown that with the deformed Lorentz symmetry, the previously mentioned problems will be naturally avoided.
The theory is self consistent even superluminal effects exist. Here we demonstrate this by showing how an example problem discussed in Ref.\cite{bi} can be resolved with deformed Lorentz symmetry, the problem of $\pi^+ \to \mu^+ \nu_\mu$.
The quantity $F_\nu(p_\nu, m_\nu) E_\nu$ plays the role of energy in the usual Lorentz symmetry theory.  The
conservations of energy and momentum in this case are given by
\begin{eqnarray}
\vec p_\pi = \vec p_\mu + \vec p_\nu\;, E_\pi = E_\mu + F(p,m) E_\nu\;.
\end{eqnarray}
The above conservation law for energy looks dramatically different than the usual one. However, if one replaces
$F(p_\nu,m_\nu) E_\nu = \sqrt{\vec p^2_\nu + m_\nu^2}$ using eq.(\ref{gg}), the energy and momentum conservation equations become
\begin{eqnarray}
\vec p_\pi = \vec p_\mu + \vec p_\nu\;, \sqrt{\vec p_\pi^2 + m^2_\pi} =\sqrt{\vec p^2_\mu + m^2_\mu} + \sqrt{\vec p^2_\nu + m^2_\nu}\;.
\end{eqnarray}
The above equations are the usual energy and momentum conservation laws expressed in terms of momenta. $\pi^+ \to \mu^+ \nu_\mu$ is not forbidden to decay in any inertial frame in this theory.

At present stage we do not have a underlying theory to provide the deformed Lorentz symmetry to dictate what are the conserved quantities, one can therefore, without loss of generality, translate any given dispersion relation into the above form and choose conserved quantities accordingly. One can also choose to re-write the general dispersion relation in the form $E^2 - G^2(p,m) p^2 = m^2$. F and G are related by $G^2 = 1+ (1-F^2)E^2/p^2$. In this case, the conserved quantities are $E$ and $G(E,m)\vec p$. In the following we will work with the form of dispersion relation in eq.(\ref{gg}).

A straightforward calculation for the velocity $v$ of a particle satisfying the dispersion relation above gives
\begin{eqnarray}
v = {dE\over dp} &=&  {p\over E} \left ({1\over F^2(p,m)} - 2 {E^2\over F(p,m)} {dF(p,m)\over dp^2}\right )\nonumber\\
&=& {p\over \sqrt{p^2+m^2}} {1\over F(p,m)} + \sqrt{p^2+m^2} {d\over dp} {1\over F(p,m)}\;.\label{vv}
\end{eqnarray}
Since the neutrino mass $m$ is small which can be neglected for practical discussions. In this case the relation between
superluminality $\delta v$ and $F = F(p,0)$ is simple with
\begin{eqnarray}
\delta v = {1\over F(p,m)} + p {d\over dp} {1\over F} -1\;.\label{vvv}
\end{eqnarray}

Whether the theory allows superluminal velocity depends on the specific form of $F(p,m)$.
For example, the dispersion relation\cite{deform1}
\begin{eqnarray}
E^2 - p^2 = m^2 + 2 E^2p^2/M^2\;,\label{dp1}
\end{eqnarray}
gives  $F^2 = 1-2p^2/M^2$, and a superluminal velocity $v$ with
\begin{eqnarray}
v = {p\over \sqrt{(p^2+m^2)}}{1+m^2/M^2\over (1-2p^2/M^2)^{3/2}} \approx 1- {m^2\over 2 p^2} + {3p^2\over M^2}\;.
\end{eqnarray}
where $m<<M$ has been assumed for the approximation in the last step above.

If neutrinos satisfy deformed Lorentz symmetry, there is a chance that they have superluminal velocity without the problems given in Refs.\cite{cg, bi}.
What form of $F(p,m)$ the neutrinos take is not known unless an underlying theory is known. Unfortunately at this stage, the study of deformed Lorentz symmetry
is not yet ready to provide such insight. We therefore will consider, at phenomenological level, dispersion relations which can provide large superluminal
velocity indicated by the OPERA data and at the same time satisfy the constraints from other neutrino data.

In general $F(p,m)$ can depend on neutrino spices, that is $F_{i = \nu_e,\nu_\mu,\nu_\tau}(p,m_i)$ may be different for different neutrinos. However,
with a small difference in $F_i$ between different ``i'', the coherence of neutrino mixing will be destroyed leading to no oscillation between neutrinos. For this reason,
we consider that the form of $F$ is universal for all neutrinos. With this assumption, one then is required to explain the energy dependence of the superlumnality
for electron neutrino $\nu_e$ and muon neutrino $\nu_\mu$. In the energy range of the OPERA data, no obvious energy dependence of the superluminality is observed. At
energies 13.8 GeV and 40.7 GeV, the associated early arrival times are $\delta t_1 = (54.7\pm 18.4(stat.)^{+7.3}_{-6.9}(sys.)) ns$
and $\delta t_2 = (68.1\pm 19.1(stat.)^{+7.3}_{-6.9}(sys.)) ns$ which implies a superluminality $\delta v$ of order a few times of $10^{-5}$ and the
averaged $\delta v = (2.37\pm 0.32(stat.)^{+0.34}_{-0.24})\times 10^{-5}$.
However, observational data from SN1987a constrain the superluminality $\delta v$ of electron neutrino to be less than $2 \times 10^{-9}$ with an energy of 10 MeV or so.
This can be taken as an indication of energy dependence since we are assuming universal dispersion relation. There are also constraints on neutrino superluminality
at higher energies up to 200 GeV.
Between MINOS\cite{minos} energy of a few GeV and the FERMILAB 1979\cite{fermilab} energy of up to 200 GeV, the energy dependence is not obvious.
The experimental data are depicted in Fig. 1. A viable dispersion relation should be able to fit the
OPERA, SN1987a and other data simultaneously.

\begin{figure}
\includegraphics[width=10cm]{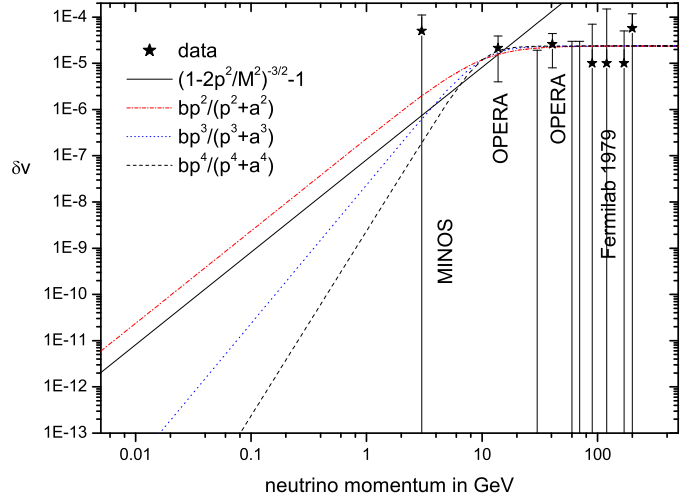}
  \caption{Constraints on neutrino superluminalities from data and predictions from dispersion relations discussed in the text for $M = 6\times 10^{3}$ GeV, $a = 10$ GeV and $b = 2.37\times 10^{-5}$.}
\end{figure}

The dispersion relation in eq.(\ref{dp1}) gives $\delta v = 1/(1-2 p^2/M^2)^{3/2}$.
We now check if this can fit the data. This dispersion relation has only one unknown parameter $M$.
To have $\delta v = 2.37\times 10^{-5}$ with energy equal to 17 GeV, the mass parameter $M$ is determined to be about $6\times 10^3$ GeV. The predictions for superluminality
at other energies of this dispersion relation is plotted in Fig. 1. It can be seen that the SN1987a limit on neutrino superluminality
is satisfied. However, at higher energies, this dispersion related produces a too large superluminality than that allowed from FERMILAB 1979 data. This
dispersion relation cannot be a viable
possible candidate.

Other dispersion relation is needed to fit the data.
To find a realistic superluminality profile, we note that there is no obvious energy dependence between the energy (around 17 GeV)
of the OPERA data and higher energies (up to 200 GeV) from FERMILAB 1979 data with a superluminality of order $10^{-5}$ or lower,
but at the energies of the SN1987a neutrino, a few MeV, the profile drops down to a upper limit less than $10^{-9}$. This suggest a form for $\delta v$
\begin{eqnarray}
\delta v = b {p^n\over a^n + p^n}\;,\label{super}
\end{eqnarray}
where a and b are constants. For large enough $p$, $\delta v$ approaches a constant b, and for small enough $p$, $\delta v \approx b p^n/a^n$ which goes to zero quickly for a positive n.

In Fig. 1, we plot the superluminaluty $\delta  v$ for n =2, 3 and 4. We agree with the observation in ref.\cite{strumia} that $n=3$ can fit data well. We use $a = 10$ GeV and $b= 2.37\times 10^{-5}$.
In fact the class of solutions in eq.(\ref{super}) can all be consistent with present data. With a larger n, it is much easier to have a small superluminality for the SN1987a data
after fitting the superluminality for the OPERA data and FERMILAB 1979 data.

If the energy profile of the superlominality $\delta v = P(p)$ is known, one can then use eq.(\ref{vvv}) to solve for $F$
and therefore the dispersion relation for neutrinos with negligibly small mass $m \approx 0$. We have
\begin{eqnarray}
{1\over F} +p  {d\over dp} {1\over F} = 1+ P(p)\;.
\end{eqnarray}
The solution for $1/F$ is given by
\begin{eqnarray}
{1\over F} = {1\over p} \left (\int^p (1+ P(p')) dp' + c\right )\;,
\end{eqnarray}
where $c$ is an integration constant set by the condition $F(0) = 1$.

For a known value of n for $\delta v$ in eq.(\ref{super}), one can  obtain the dispersion relation factor $F$ using the above equation. We write down the general solution for $F$ with integer n, $F_n(p)$,
\begin{eqnarray}
{1\over F_n(p)} &=& 1 + b \nonumber\\
 &-&{a b\over p} \left \{\begin{array}{l}
{2\over n} \sum^{n/2-1}_{k=0} [Q_k\sin({(2k+1)\pi\over n}) - P_k \cos({(2k+1)\pi\over n})]\;,\;\;\mbox{n = even}\\
\\
{1\over n} \ln((1+{p\over a}) + {2\over n} \sum^{(n-3)/2}_{k=0} [Q_k\sin({(2k+1)\pi\over n}) - P_k \cos({(2k+1)\pi\over n})] - C\;,\;\;\mbox{n = odd}\;.
\end{array}
\right .\;,
\end{eqnarray}
where
\begin{eqnarray}
P_k &=& {1\over 2} \ln[{p^2\over a^2} - 2{p\over a} \cos({(2k+1)\pi\over n}) + 1]\;,\nonumber\\
Q_k &=& \arctan\left({{p\over a} - \cos({(2k+1)\pi\over n})\over \sin({(2k+1)\pi\over n})}\right )\;,\nonumber\\
C &=& \sum^{(n-3)/2}_{k = 0}({(2k+1)\pi\over n} - {\pi\over 2})(\sin{(2k+1)\pi\over n})\;.
\end{eqnarray}

For $n = 2$ and $n = 3$, we have
\begin{eqnarray}
&&F_2(p) = {p/a\over (1+b)p/a - d \arctan(p/a)}\;,\\
&&F_3(p) = {6\sqrt{3} p/a\over 6\sqrt{3}(1+b)p/a - b (\pi + 6\arctan[(2p-a)/\sqrt{3}a] + \sqrt{3}\ln[(a+p)^2/(a^2-ap+p^2)])}\;.\nonumber
\end{eqnarray}

We therefore have found a class of deformed Lorentz symmetry which can fit the current data on neutrino superluminality $\delta v$. As
have been shown earlier, with the function $F(p,m)$ known one also knows how the energy- momentum conservation laws are modified.
The conserved quantities are $F(p,m) E$ and $\vec p$. The momentum conservation law is not changed compared to the usual Lorentz symmetry.
The energy conservation law is modified. In principle one can test the $F(p,m) E$ law against the usual $E$ conservation law in the usual Lorentz symmetry.
However, the deviation of $F(p,m)$ from 1 in the dispersion relation is very small, at most of order a few times of $10^{-5}$ as shown in Fig.2 for n =2, 3 and 4 cases.
It is, therefore, difficult to test these modified energy conservation laws at present. The deviation at high energies for $F^2 = 1-2p^2/M^2$ case can be larger. Unfortunately, this
case also produce too large a superluminality at energy larger than OPERA neutrino energy and therefore it cannot be a  viable candidate.

\begin{figure}
  \includegraphics[width=10cm]{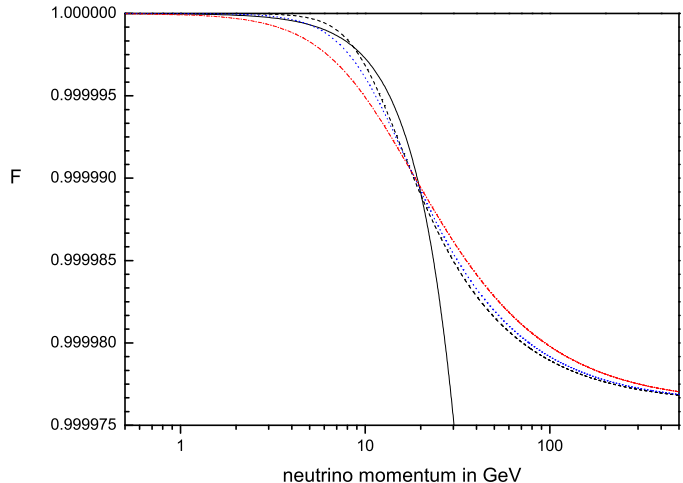}
  \caption{$F(p)$ as functions for $\delta v = 1/(1-2 p^2/M^2)^{3/2}$ and $\delta v = b p^n/(p^n+a^n)$ with $n =2, 3, 4$.}
\end{figure}

We have studied dispersion relations which can be interpreted as from deformed Lorentz symmetry. We find that it is possible
to have dispersion relations which can explain data on neutrino superluminality data. We have shown that once the superluminality
$\delta v$ as a function of momentum is known, one can determine the corresponding dispersion relation and therefore the conservation law corresponding to the energy
conservation in a theory with the usual Lorentz symmetry. The energy conservation law is
modified which in principle can be used to test the models. The modifications are, however, small and cannot be tested at the present.
Future experiments which can measure neutrino energy to high precision may provide useful information about deformed Lorentz symmetry.

\acknowledgments
This work was supported in part by NSC and NCTS of ROC, and SJTU 985 grant and NNSF grant of PRC.


\end{document}